\documentclass[10pt,letterpaper,twocolumn]{article} 

\usepackage{ol2}
\usepackage[draft,implicit=false]{hyperref}
\usepackage{amsmath}

\begin{document}

\twocolumn[

\title{Phase Singularity Diffusion}
\author{Xiaojun Cheng,$^{1,2*}$ Yitzchak Lockerman,$^1$ and Azriel Z. Genack$^{1,2}$}
\address{
$^1$Department of Physics, Queens College,
The City University of New York, Queens, NY 11367, USA\\
$^2$The Graduate Center, The City University of New York, New York, NY 10016 USA\\
$^*$Corresponding author: xcheng@qc.cuny.edu}

\begin{abstract}
We follow the trajectories of phase singularities at nulls of intensity in the speckle pattern of waves transmitted through random media as the frequency of the incident radiation is scanned in microwave experiments and numerical simulations. Phase singularities are observed to diffuse with a linear increase of the square displacement $\langle R^2\rangle$ with frequency shift. The product of the diffusion coefficient of phase singularities in the transmitted speckle pattern and the photon diffusion coefficient through the random medium is proportional to the square of the effective sample length. This provides the photon diffusion coefficient and a method for characterizing the motion of dynamic material systems.
\end{abstract}

\ocis{(030.6600) Statistical optics; (290.4210) Multiple scattering; (290.7050) Turbid media.}
 ]

\noindent The decorrelation in time of intensity of a monochromatic laser beam scattered from dynamic single- and multiple-scattering samples can provide the size distribution of macromolecules and biological cells \cite{Pecora1972}. Quasi-elastic light scattering can also be used to characterize the motion of the constituents of dense complex media and to image tissue based on the temporal or spatial variations of speckle patterns \cite{Stern1975,Fercher1981}. Variations of the photon correlation time across an inhomogeneous sample can also be determined from measurements of the contrast in the speckle pattern over time intervals comparable to the decorrelation time \cite{Boas2010}. The lower the contrast of the time averaged speckle pattern, the more rapidly the intensity in that region fluctuates as a result of motion within the sample. This has permitted a precise imaging of superficial blood vessels in the brain \cite{Boas2001}. In static samples, the correlation function of the field with frequency shift provides the time evolution of pulses reflected from or transmitted through random media and the diffusion coefficient for light or other electromagnetic radiation \cite{Genack1990}. 
 
Because the speckle pattern is generic, the statistics of its structure do not change as the speckle pattern evolves with changes in the sample configuration or incident frequency \cite{Soskin1997,Takeda2005,Berry2000,Sheng2007,Sheng2010}. It is thus possible in principle to characterize the interaction of the wave through the sample by following the displacement of specific features of the speckle pattern in time or frequency. The most distinctive features of a speckle pattern are the phase singularities at intensity nulls of the speckle pattern for each polarization component of the wave \cite{Freund1994,Takeda2005,Kirkpatrick2012}. Since both the in- and out-of-phase components of a single polarization of the field vanish at an intensity null, phase is ill-defined there. The phase is singular with a discontinuity of $\pi$ rad along any line passing though the center of the electromagnetic vortex centered on the singularity. The phase change in a complete counterclockwise circuit around the phase singularity is $\pm2\pi$ rad, thus they are easy to locate in speckle patterns. This is associated with a topological charge of $\pm 1$ \cite{Freund1994}. Higher order vortices do not appear to arise in speckle patterns of waves transmitted through random media\cite{Sheng2007,Takeda2005}. Phase singularities of opposite signs are created or annihilated in pairs in a process in which charge is conserved. Though much is known about the structure of speckle patterns, their motion has been little studied. The velocity statistics of singularities has been calculated \cite{Berry2000} and measured \cite{Sheng2007,Sheng2010} and the variance of singularity velocity has been shown to be a measure of the impact of weak localization \cite{Sheng2007}. However, the statistics of motion over stretches of time in dynamic samples or over frequency change in static samples have not been characterized.

In this Letter, we utilize microwave spectral measurements and computer simulations in the frequency and time domains to demonstrate that singularity motion in the speckle pattern transmitted through a scattering medium is diffusive. The vanishing of the velocity correlation function with frequency $\Delta\nu$  or time $\Delta t$  leads to a linear increase in the average of the square of the displacement of phase singularities $\langle R^2\rangle$. This yields the diffusion coefficient of phase singularities $D_{\mathrm{s}}'$, which is normalized by the scale of the speckle pattern. The photon diffusion coefficient is shown to be proportional to $L_{\mathrm{eff}}^2/D_{\mathrm{s}}'$, where the effective sample length $L_{\mathrm{eff}}$ incorporates the impact of internal reflection on both sides of the sample interfaces. Measurements of $D_{\mathrm{s}}'$ reflect the motion within dynamic media and may provide a useful means for mapping internal motion of a sample for imaging and diagnostic applications. 

Universal properties of the statistics of singularity trajectories are studied in microwave measurements. The sample is composed of alumina spheres with diameter 0.95 cm and refractive index 3.14 embedded in Styrofoam shells, which gives an alumina volume fraction of 0.068. These spheres are contained inside a copper tube with diameter 7.2 cm and length 61 cm. Speckle patterns are recorded at the output on a 1-mm grid. The full field is recovered with use of the Whittaker-Shannon two-dimensional sampling theorem \cite{Sampling}. The frequency is scanned from 14.7 to 15.7 GHz in 1600 frequency steps for 40 random realizations of the sample. New sample configurations are created by rotating and vibrating the sample tube.  

\begin{figure}[htbp]
\centerline{\includegraphics[width=.8\columnwidth]{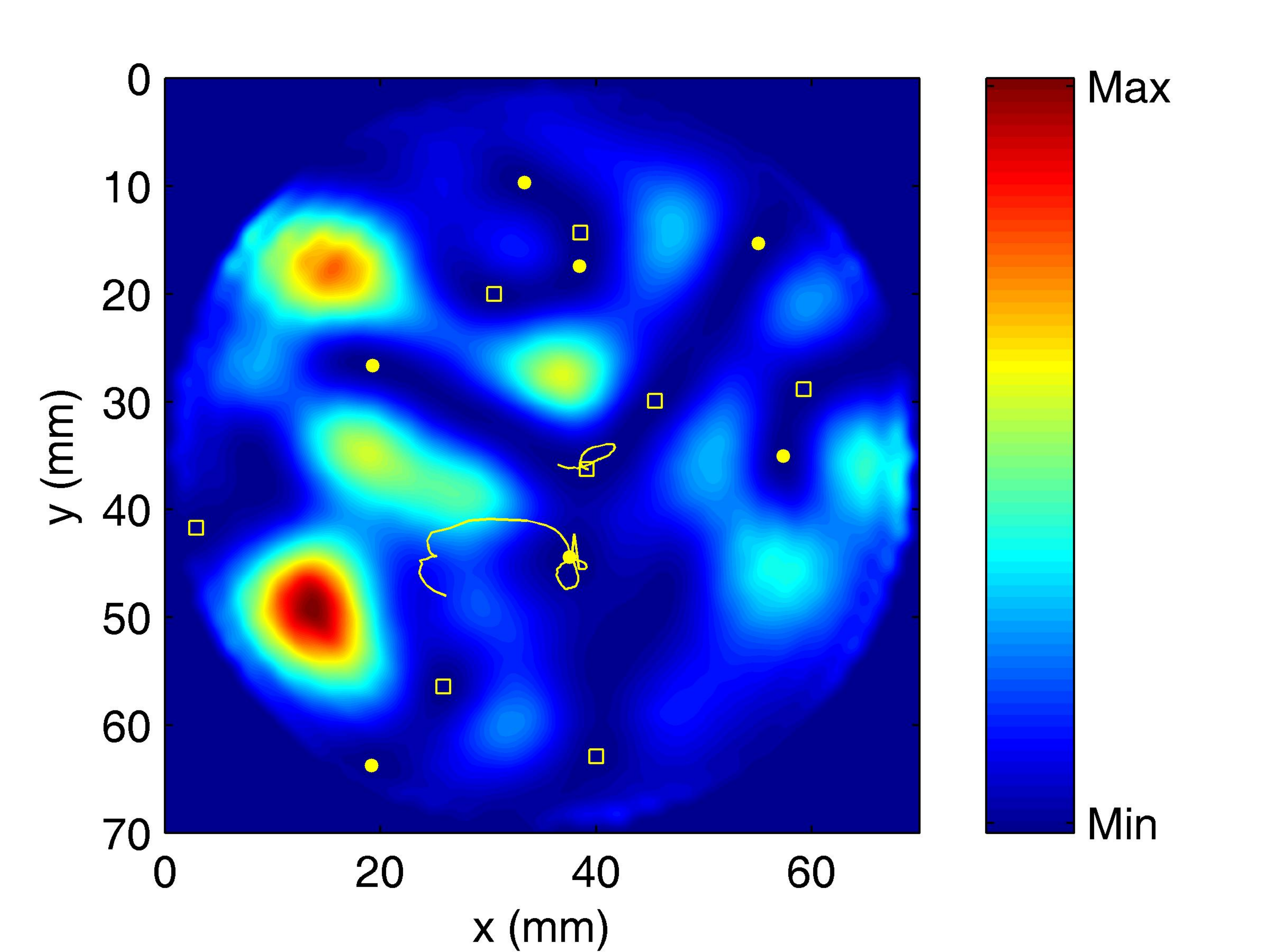}}
\caption{Singularities in the speckle pattern in microwave experiments. Singularities with positive and negative charges are represented by yellow dots and yellow squares respectively. We show trajectories of two singularities as examples. Note that since singularities are annihilated after a typical decaying frequency, as will be shown in Fig. \ref{fig:decaySing}, the comparably longer trajectory as in this figure is a relatively rare event.}
 \label{fig:Speckle}
\end{figure}

\begin{figure}[htbp]
\centerline{\includegraphics[width=.8\columnwidth]{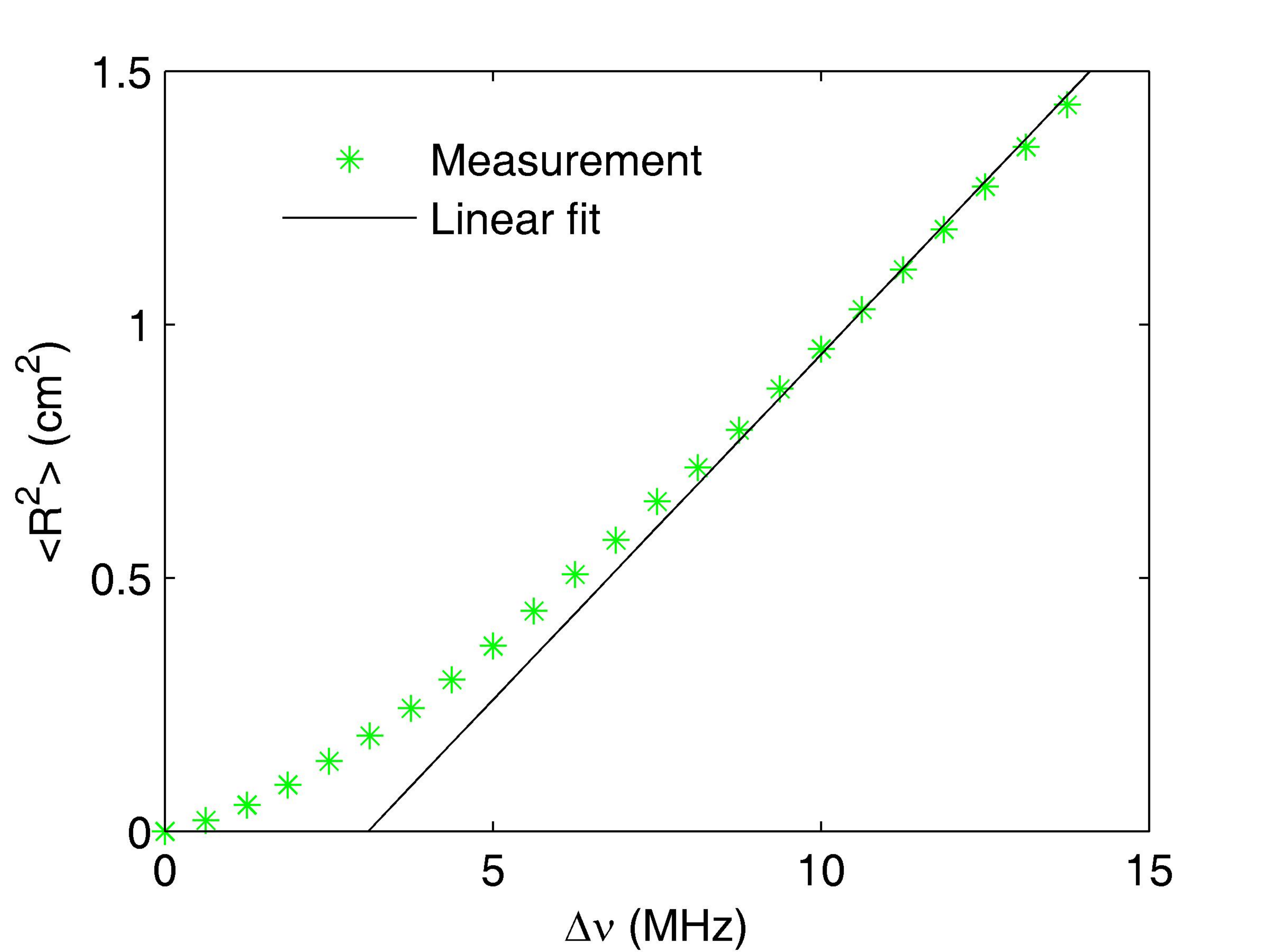}}
\caption{Average of the square of singularity displacement with frequency shift. Slope of $\langle R^2\rangle$ approaches a constant.}
 \label{fig:diffusiveR2}
\end{figure}

We follow the trajectories of singularities as frequency is stepped. An example of singularities in a speckle pattern with the trajectories of two singularities is shown in Fig. \ref{fig:Speckle} and the average square displacement $\langle R^2\rangle$ vs. frequency shift $\Delta\nu$ is shown in Fig.  \ref{fig:diffusiveR2}. $\langle R^2\rangle$  increases linearly after a frequency shift of approximately 10 MHz.

\begin{figure}[htbp]
\centerline{\includegraphics[width=.8\columnwidth]{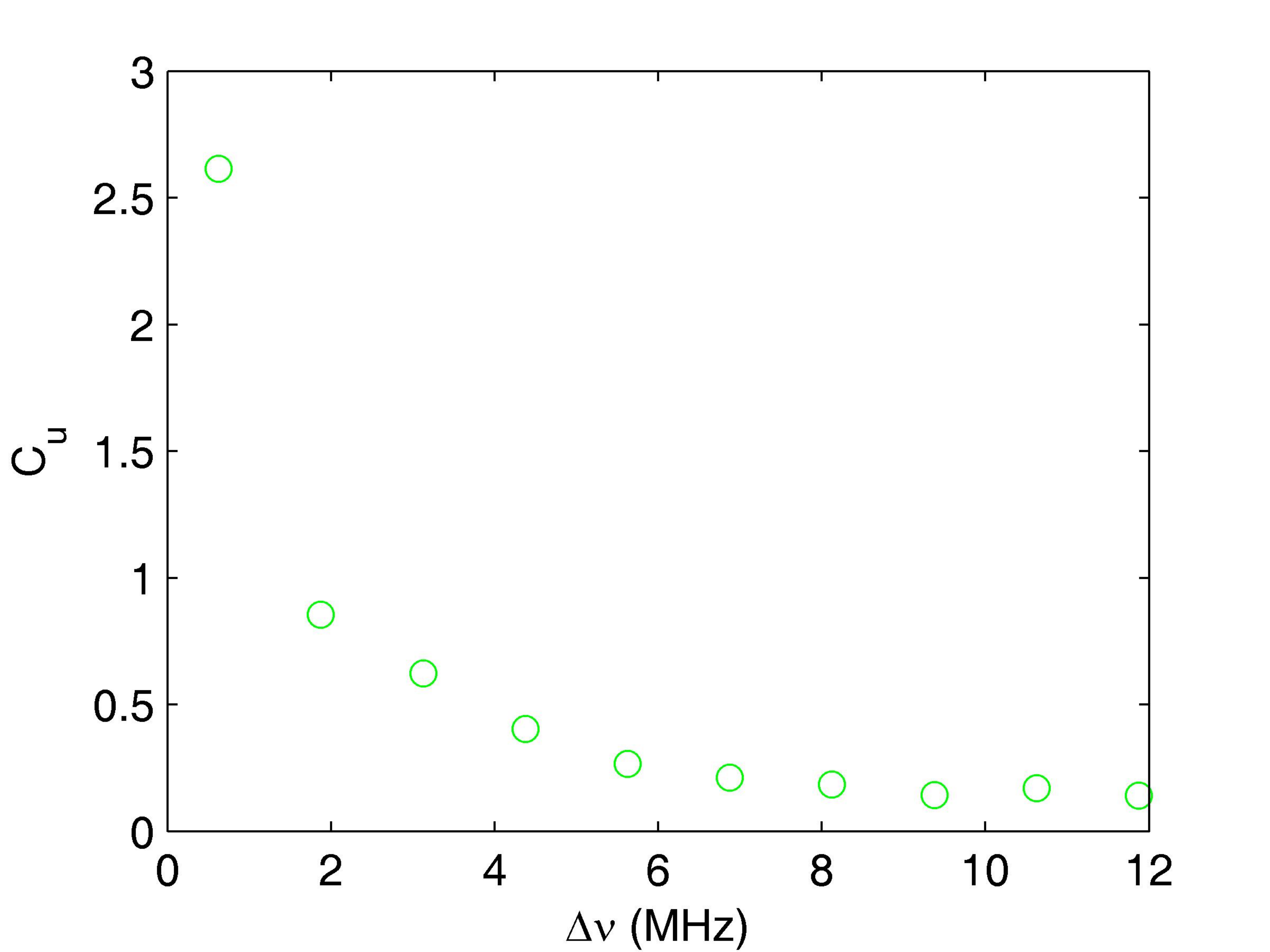}}
\caption{Measurement of the velocity auto-correlation function with frequency shift.}
 \label{fig:cu}
\end{figure}

The average of the square displacement is given by the velocity correlation function \cite{Kubo}, 
\begin{equation}
\langle R^2(t)\rangle=\langle (\vec{r}(t)-\vec{r}(0))^2\rangle=\int^t_0dt'\int^t_0dt''\langle \vec{u}(t')\vec{u}(t'')\rangle. 
\end{equation}
Here $\vec{r}$ is the position of a singularity. The variable $t$ denoting time can be replaced by $\nu$ when considering the variation with frequency shift studied experimentally. The velocity auto-correlation function with frequency shift $C_u(\Delta\nu)=\langle(\vec{u}(\nu)\cdot\vec{u}(\nu+\Delta\nu))\rangle/\langle u\rangle^2$ shown in Fig. \ref{fig:cu} decays rapidly. The loss of correlation indicates that the motion of a phase singularity is random. The correlation function in the limit of vanishing frequency step $\Delta\nu=0$, $C_u(0)=\langle u^2\rangle/\langle u\rangle^2$ diverges. This can be calculated with the probability distribution of velocities of phase singularities $P(\bar{u})=\frac{8\pi^2\bar{u}}{(\pi^2\bar{u}^2+4)^2}$,where $\bar{u}=u/\langle u\rangle$ \cite{Berry2000,Sheng2010}. 

\begin{figure}[htbp]
\centerline{\includegraphics[width=.8\columnwidth]{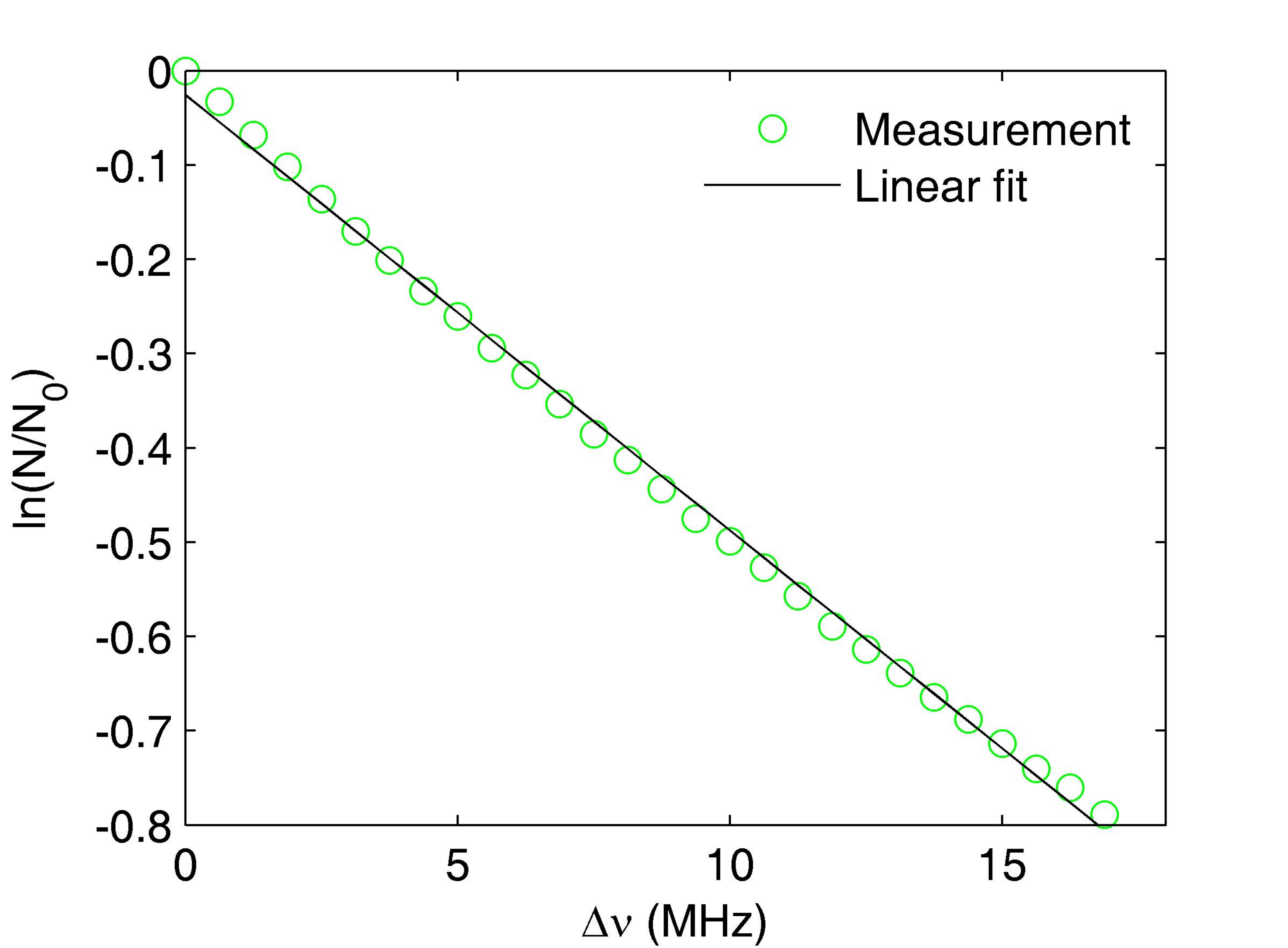}}
\caption{Decay of the number $N$ of phase singularities with frequency shift in microwave experiments. $N_0$  is the initial number of singularities. $N/N_0$ falls to $1/e$ in a frequency shift $\Delta\nu_\mathrm{N}=$20.4 MHz.}
 \label{fig:decaySing}
\end{figure}

The number of phase singularities that survive annihilation by oppositely charged singularities as the frequency is shifted is shown to fall exponentially in Fig. \ref{fig:decaySing}. Here we track the singularities over a limited frequency range with initial positions near the center of the output so that singularities will not disappear over the edge of the speckle pattern. The frequency $\Delta \nu_\mathrm{N}$ in which the number of singularities present at an initial frequency is reduced by a factor of $e$ provides a characteristic frequency shift in which the speckle pattern changes, which is obtained without a measurement of the field. We expect that when the displacement of singularities $R$ is normalized by the inverse of the square root of the singularity density to give $R'$ and the frequency shift is given in units of $\Delta \nu_\mathrm{N}$, $\Delta \nu' = {\Delta \nu}/\Delta \nu_\mathrm{N} $, the variation of $\langle R'^2\rangle$ with $\Delta \nu'$ will be universal.  It is important to normalize $R$ by the relevant scale of the speckle pattern so the results do not depend upon the degree of enlargement of a speckle pattern and reflect only the evolution of the generic structure of the pattern. Deviations might arise at early times or small frequency shifts due to differences in angular distribution of waves. Singularities also move more rapidly when they are just created or annihilated and this non-ballistic motion is quite complex, which is out of the scope of this paper.


\begin{figure}[htbp]
\centerline{\includegraphics[width=.8\columnwidth]{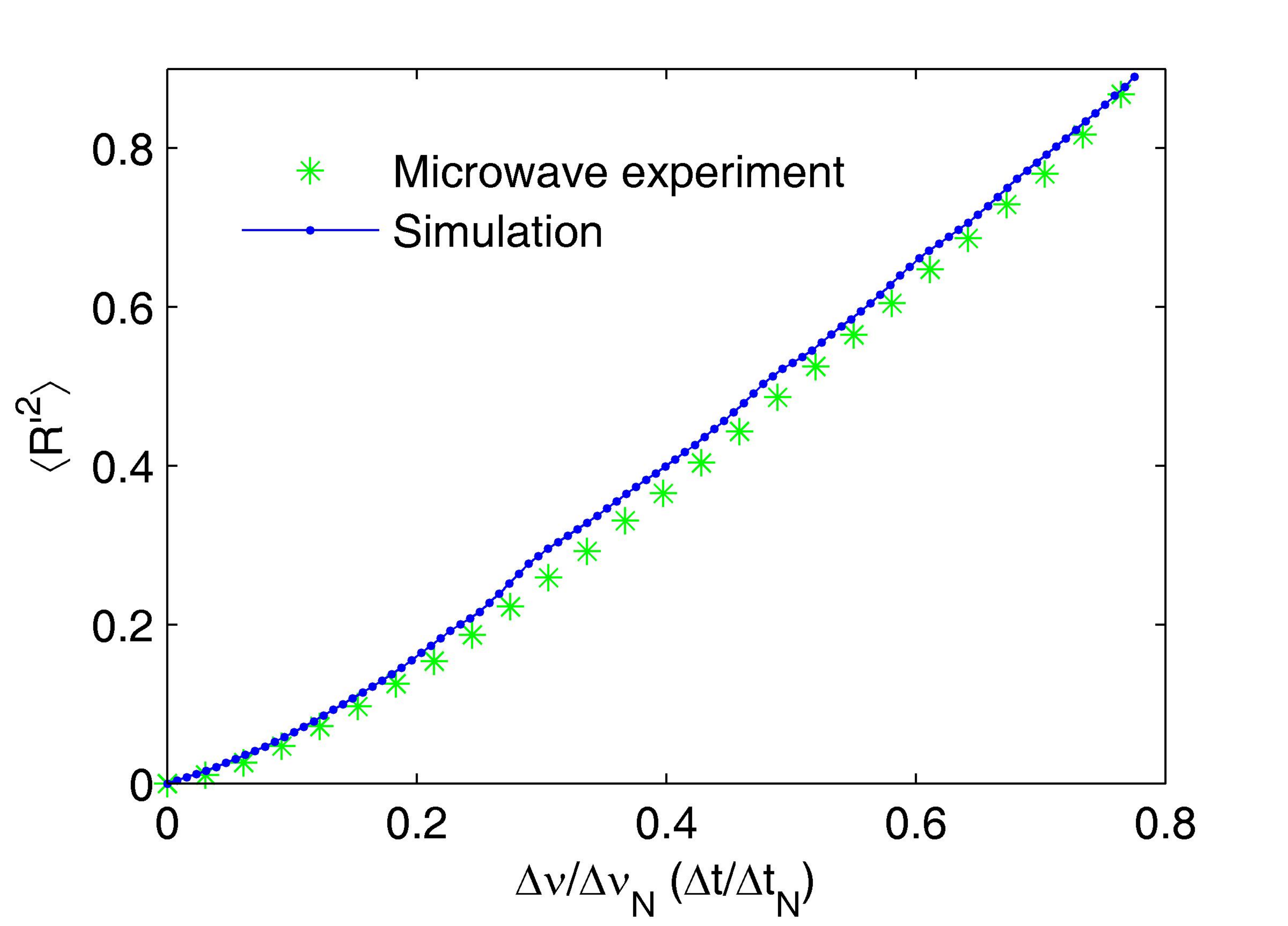}}
\caption{Comparison of the square displacement of phase singularities in microwave experiments in the frequency domain and simulations in the time domain. The displacement, frequency shift and time are normalized in terms of the singularity density and survival as explained in the text.}
 \label{fig:CompareES}
\end{figure}

We compare results of frequency domain experiments to random wave simulations in the time domain. Speckle patterns in a plane are generated in the simulation by the superposition of 300 randomly phased waves in the plane, $E(x,y,t)=\sum_iA_i\mathrm{exp}[ik_{ix}x+k_{iy}y-\omega_i t]$, where $k_{ix}$, $k_{iy}$, $k_{iz}$ and $A_i$ are randomly drawn from Gaussian distributions with zero mean and unit variance and $\omega_i=c\sqrt{k_{ix}^2+k_{iy}^2+k_{iz}^2}$. We find good agreement in Fig. \ref{fig:CompareES} between measurements in the plots of of $R'$ vs. $\Delta \nu'$ and simulations of $R'$ vs. $t' \equiv t/t _\mathrm{N}$. Within $\nu_\mathrm{N}$ or $t_\mathrm{N}$, singularities will migrate a distance about the average spacing of singularities, or $R'\sim 1$. With the normalized scale, the diffusion behavior is universal, independent of the values of experiment and simulation parameters.

\begin{figure}[htbp]
\centerline{\includegraphics[width=.8\columnwidth]{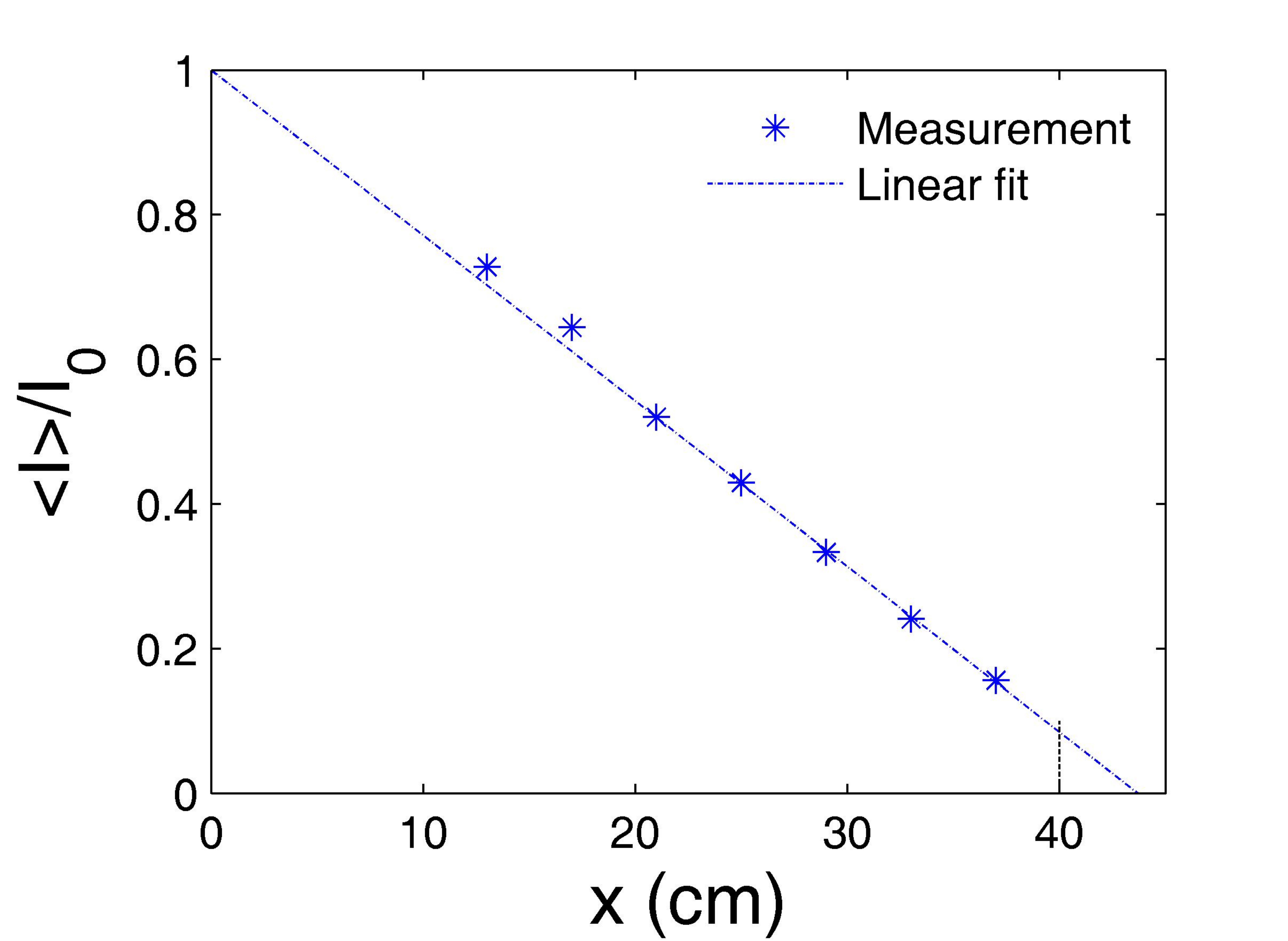}}
\caption{Linear decay of intensity averaged over the cross section with depth $x$ into the sample $\langle I(x)\rangle$ for sample with $L$=40 cm determined from FDTD simulations. 
$I_0$ is the value to which the average intensity within the sample extrapolates at the input surface $x$=0. The intensity profile within the sample extrapolates to zero at a length $z_{\mathrm{b}}$=3.85 cm beyond the output surface.}
 \label{fig:IntensityProfile}
\end{figure}

The displacement of phase singularities can be characterized in terms of a two-dimensional diffusion coefficient with displacement normalized to the scale of the speckle pattern $ D'_\mathrm{s}$, which is obtained from the diffusion of singularities relative to the scale of the speckle pattern, and the photon diffusion coefficient through the random medium $D_\mathrm{p}$. The diffusion of singularities in two dimensions gives
\begin{equation}
\langle R'^2\rangle=4D'_\mathrm{s}\Delta\nu. 
\end{equation}
The time of flight distribution of waves transmitted through a random medium is the Fourier transform of the field correlation function. As a result the field correlation frequency is proportional to the inverse of the width of the time of flight distribution \cite{Genack1990}. For diffusive waves this gives the correlation frequency, $\delta\nu\sim\Delta \nu_\mathrm{N}\sim D_{\mathrm{p}}/L_{\mathrm{eff}}^2$. Here, $L_{\mathrm{eff}}=L+2z_\mathrm{b}$ is the effective sample thickness and $z_\mathrm{b}$ is is the length beyond the sample in which the linear falloff of intensity inside the sample extrapolates to zero. We expect that the field is decorrelated once singularities are displaced on average by a distance comparable to the size of a speckle spot so that $R' \sim 1$ when $\Delta \nu=\Delta \nu_\mathrm{N}$. Then at the frequency shift, $\Delta \nu=\Delta \nu_\mathrm{N}$, we have $4D'_\mathrm{s}(D_{\mathrm{p}}/ L_{\mathrm{eff}}^2)=C$, where $C$ is a constant of order unity. Thus $D_\mathrm{p}$ can be obtained from observations of the motion of intensity nulls in the speckle pattern and should be inversely proportional to $D'_\mathrm{s}$ via,
\begin{equation}\label{eq:dpds}
D_{\mathrm{p}}=\frac{CL_{\mathrm{eff}}^2}{4D'_\mathrm{s}}.
\end{equation} 
To confirm Eq. \ref{eq:dpds}, we carried out finite difference time domain (FDTD) simulations on a structure of spheres with the same dielectric function and diameter as in the experiments using the Omnisim program from Photon Design, Inc. The frequency range in the simulations is 14.49-14.86 GHz over which the wave is diffusive for the lengths and sphere densities studied. The simulations give the intensity profile inside the system as shown shown in Fig. \ref{fig:IntensityProfile}. The intensity extrapolates to zero at a distance $z_\mathrm{b}=3.85$ cm beyond the output face of the sample with density of 0.068 and sample length $L=40$ cm. The intensity at a length is averaged over 1600 points measured on the cross section for 50 configurations. We did not include data points close to the input because the wave is then not totally randomized \cite{ShengBook2006}.
 
\begin{figure}[htbp]
\centerline{\includegraphics[width=.8\columnwidth]{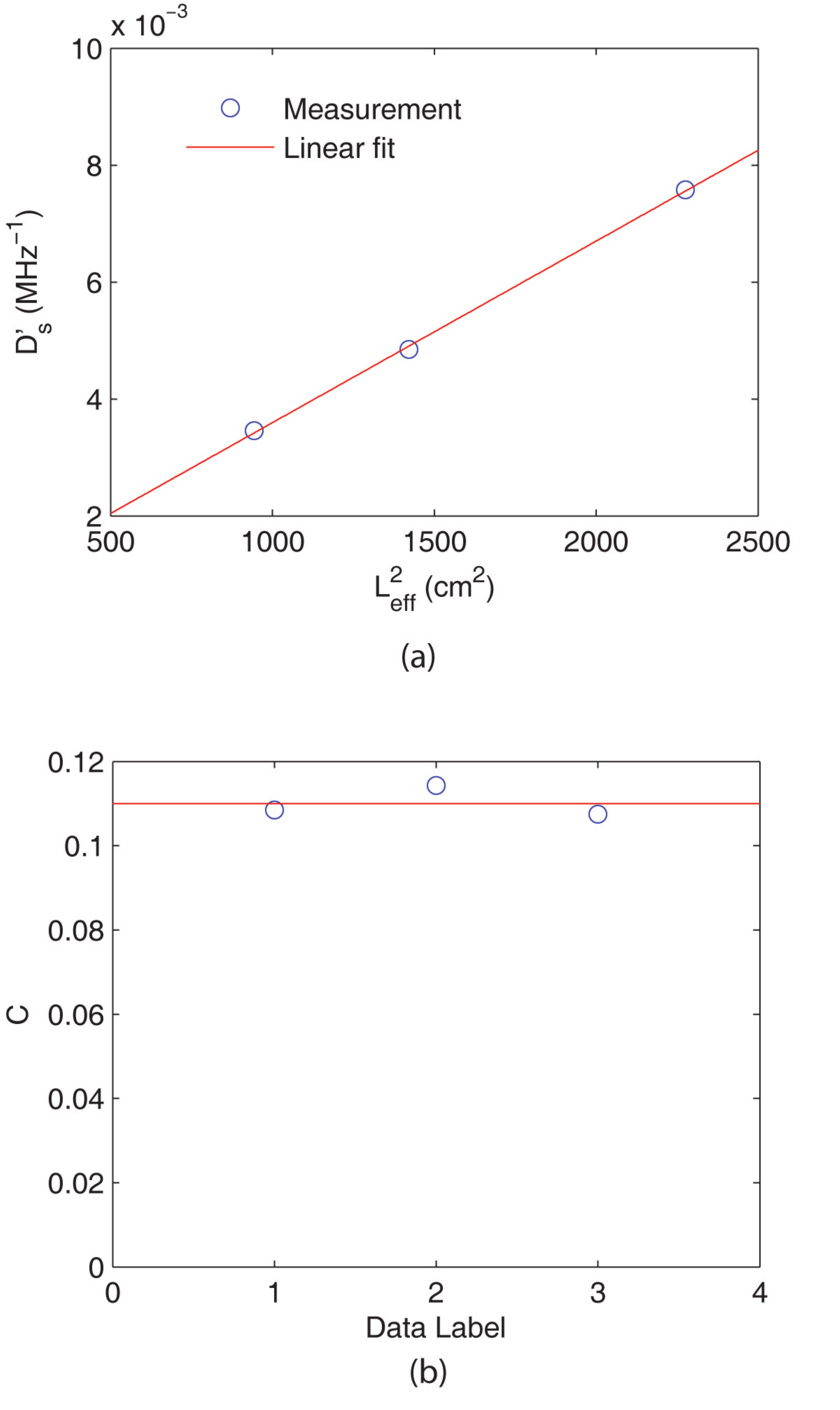}}
\caption{The relation between the diffusion coefficient of singularities in the output plane $D'_\mathrm{s}$, the diffusion coefficient of photons through the sample $D_\mathrm{p}$, and the effective sample length $L_{\mathrm{eff}}$ in FDTD simulations. (a) Results are shown for three values of $L$ with the same structure and so the same value of $D_\mathrm{p}$. $D'_\mathrm{s}$ is proportional to $L_{\mathrm{eff}}^2$. (b) The relationship between $D'_\mathrm{s}$ and $D_\mathrm{p}$ at $L=23$ cm for samples 1, 2 and 3 correspond to filling fractions of alumina spheres of $0.068$, $0.1$ and $0.15$. Here $C=4D'_\mathrm{s}(D_{\mathrm{p}}/ L_{\mathrm{eff}}^2)$.}
 \label{fig:Ds}
\end{figure}

Once $z_\mathrm{b}$ is determined, we obtain $L_{\mathrm{eff}}$ and can check the validity of Eq. \ref{eq:dpds}. The dependence of $D'_\mathrm{s}$ on $L_{\mathrm{eff}}$ is shown in Fig. \ref{fig:Ds}(a). $D'_\mathrm{s}$ is seen to increase linearly with $L_{\mathrm{eff}}^2$. We have also demonstrated the inverse relationship between $D'_\mathrm{s}$ and $D_\mathrm{p}$ holds in simulations for three samples with alumina filling fractions of $f=0.068, 0.1, 0.15$ in a $23 $-cm-long copper tube with all other simulation parameters the same as described above. $D_\mathrm{p}$ is obtained from the average dwell time $\tau=d\varphi/d\omega=\frac{L_{\mathrm{eff}}^2}{6D_\mathrm{p}}$ \cite{Landauer1987}. The dimensionless constant $C$ in Eq. \ref{eq:dpds} is found to be $0.11\pm 0.1$ from the simulation. 

In conclusion, we found that phase singularities in the speckle pattern of light transmitted through a random medium diffuse with frequency shift or time delay and that their diffusion coefficient is inversely proportional to the photon diffusion coefficient within the medium.  Since frequency and time are conjugate variables, analogous results are obtained in the frequency and time domains. The generic diffusion of phase singularities does not depend on the scale of the system so that results obtained in microwave measurements hold at optical wavelengths as well. Applications of this work include measurements of the photon diffusion coefficients in complex systems from radio to optical frequencies as well as the diffusion coefficients of vibrational excitation for sound, ultrasound and acoustic phonons.  The variation of singularity motion in speckle patterns of monochromatic light reflected from samples which are inhomogeneous on scales greater than a mean free path may find applications to the analysis of the structure and function in biomedical tissue. 

We thank Sheng Zhang and Jing Wang for useful discussions. This work is supported by NSF (No DMR-1207446).

\pagebreak

\end{document}